\pdfoutput=1 

\documentclass[english,prl, twocolumn]{revtex4}
\usepackage[T1]{fontenc}
\usepackage[latin9]{inputenc}
\usepackage{graphicx}
\usepackage{babel}
\usepackage{amssymb,amsmath}
\usepackage{subfigure}
\usepackage{color}
\usepackage{units}
\usepackage{comment}

\newcommand{\dif}{\mathrm{d}}
\definecolor{grey}{gray}{.35} 

\newcommand{\SI}{Supplementary Information}

\begin{document}

\title{Possible Origin of Stagnation and Variability of Earth's Biodiversity}

\author{Frank Stollmeier$^{1,2}$, Theo Geisel$^{2}$, and Jan Nagler$^{3,2}$}

\affiliation{$^1$Network Dynamics Group, Max Planck Institute for Dynamics and Self-Organization (MPI DS) G\"ottingen, Am Fassberg 17, 37077 G\"ottingen, Germany\\
$^2$Max Planck Institute for Dynamics and Self-Organization (MPI DS) G\"ottingen, and
Institute for Nonlinear Dynamics, Faculty of Physics, University of G\"ottingen, Am Fassberg 17, 37077 G\"ottingen, Germany\\
$^3$Computational Physics, IfB, ETH Zurich, Wolfgang-Pauli-Strasse 27, 8093 Zurich, Switzerland}

\begin{abstract}
The magnitude and variability of Earth's biodiversity have puzzled scientists ever since paleontologic fossil databases became available.
We identify and study a model of interdependent species where both endogenous and exogenous impacts determine the nonstationary extinction dynamics.
The framework provides an explanation for the qualitative difference of marine and continental biodiversity growth. 
In particular, the stagnation of marine biodiversity may result from a global transition from an imbalanced to a balanced state of the species dependency network.
The predictions of our framework are in agreement with paleontologic databases.

\end{abstract}

\maketitle

\begin{figure}[t]
    \includegraphics[width=5.9cm]{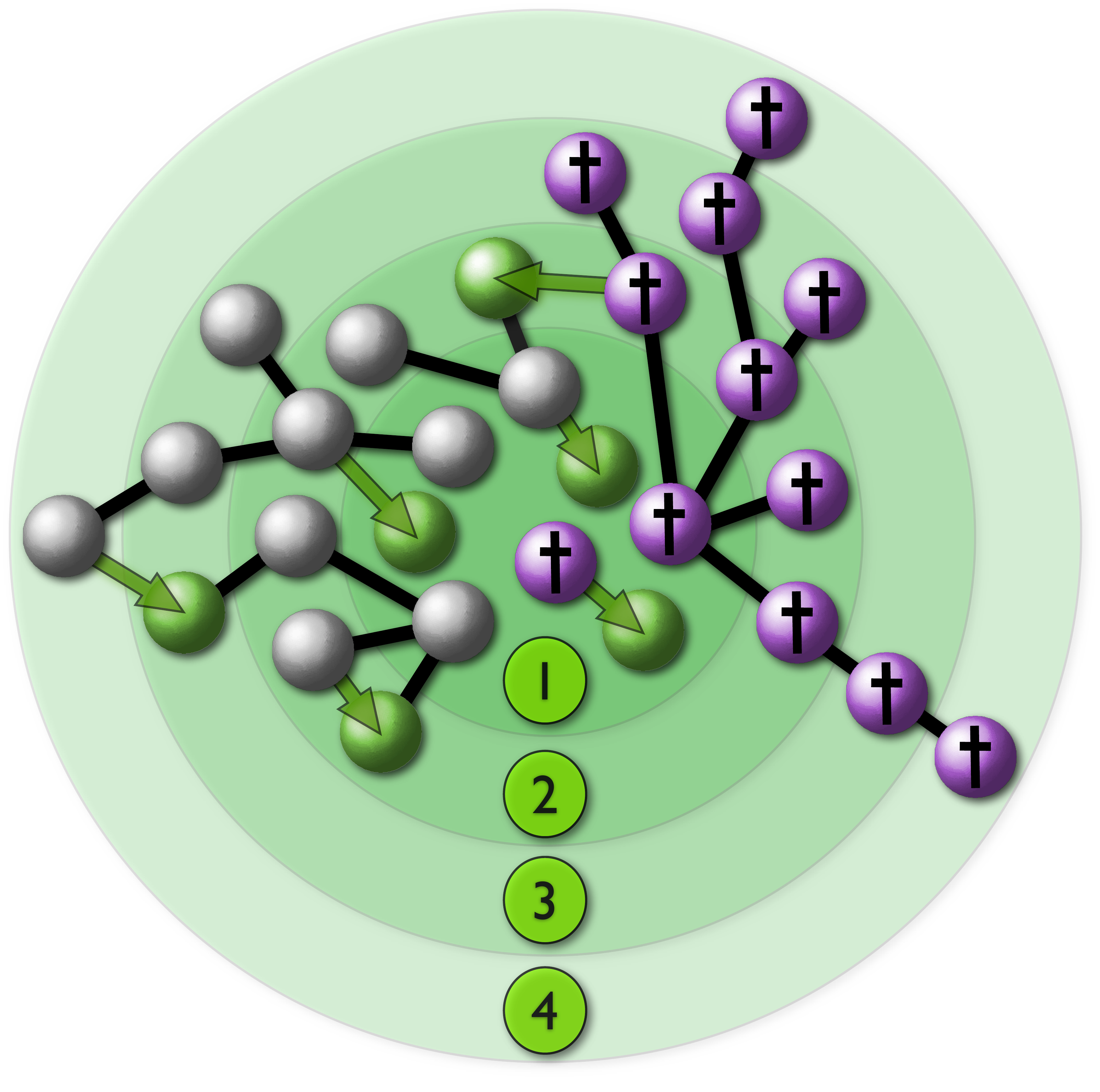}
    \caption{
    Sketch of the model. The species in purple (crosses) become extinct at level 1 by exogenous causes and at higher levels
    due to their dependence on species at lower levels. The species in green(gray) originate from existing species (arrows).}
    \label{fig:trees}
\end{figure}

Biodiversity has expanded from few species at the beginning of the Phanerozoic (541 Ma ago) to some million species today. 
Large collections of fossils
have enormously increased our understanding of the history of Earth's biodiversity.
The underlying processes governing this diversification are, however, poorly understood and 
a compelling interpretation of the fossil data remains challenging \cite{Aberhan2012}.
Whereas it is commonly accepted that continental biodiversity exhibited exponential growth \cite{Hewzulla1999,Benton2001,Morlon2010,Vermeij2010,Kalmar2010},
the growth dynamics of marine life, albeit documented in substantially larger detail, has been controversially debated.

Pioneering work on 
both the Fossil Record 2 \cite{Benton1993} and Sepkoski's compendium \cite{Sepkoski2002} suggested 
that after a first increase the total diversity remained fluctuating around a constant level for roughly 200 million years and suddenly continued to increase \cite{Raup1982,Benton1995}.
The constant level has traditionally been associated 
with the equilibrium of a logistic growth. 
In particular, Sepkoski identified three evolutionary faunas and modeled their diversity by three coupled logistic equations with model parameters fitted to the fossil data \cite{Sepkoski1984,Alroy2004}. This model  explains the emergence of a biodiversity equilibrium together with a subsequent increase.
A model combining exogenous impacts and logistic growth has been suggested by Courtillot and Gaudemer \cite{Courtillot1996}.
Their model is based on 
four time segments separated by three mass extinctions, where
each segment is described by a logistic growth process with an individual equilibrium level. 
More recently, an analysis that identifies and overcomes  
the sampling bias in previous fossil data analyses
suggests that the increase in biodiversity after the 200 million years period of stagnation may be a mere artifact \cite{Alroy2008}, and 
simple logistic growth with fluctuations around a single equilibrium level 
a sufficient model (see Supplementary Fig.\ S1).

These models are based on the assumption that new species establish and remain in presence only if they successfully compete for space or resources.
Therefore the diversity at large scales approaches an equilibrium as a result of a global logistic growth process. 
Empirical evidence for this hypothesis has been found in the fossil data \cite{Rabosky2009,Aberhan2012},
in particular the observation that the extinction and origination rates are dependent on the relative number of species \cite{Alroy2008a}.

The causes for equilibria in the diversity of marine life have been discussed controversially. 
The equilibria might be the result of an expanding diversity punctuated by extinction events \cite{Stanley2007}, 
or an artifact of the subsumption of species in higher taxonomic groups \cite{Benton2007}.
Exponential growth may display an equilibrium due to an overcompensating correction of sampling bias \cite{Vermeij2010}.
In contrast, recent studies strongly support the exponential hypothesis for continental biodiversity \cite{Benton2001,Morlon2010,Vermeij2010,Kalmar2010} suggesting 
 that either the growth dynamics of continental diversity may crucially differ from marine diversity \cite{Benton2001,Vermeij2010}, 
 or that both grow exponentially \cite{Hewzulla1999}. 

This demonstrates that arguments and empirical evidence in this debate on equilibrium and expansion are contradictory.
While there is a large body of work on logistic growth models, 
frameworks 
based on the assumption of an expanding diversity are, to our knowledge, absent. 

We present a model supporting the expansion hypothesis with few simple reasonable assumptions. 
The dynamics of this model results in exponential growth that, however, transiently slows down or is even interrupted for some time.
Specifically, while the average diversity grows exponentially, the species dependency network may develop into an unstable imbalanced state where many species depend on few. 
In our model a transiently increased extinction rate compensates the speciation rate and causes a reorganization of the network to the balanced state as being the attractor of the system.
The impact of this mechanism which results in periods of a stagnating diversity is determined by the ratio of the extinction to speciation probability.
A comparison with the fossil data suggests 
that marine and continental taxa indeed have different ratios of the extinction to the origination probability
which may explain the qualitative difference of marine and continental biodiversity growth.

\paragraph{Model}

In our model species can become extinct due to abiotic causes (random exogenous extinctions) like a changing environment,
 or are threatened by biotic causes from extinction cascades in the dependency network (endogenous extinctions). 

As a result, the size of extinction events ranges from one to all species, 
which is in agreement with the fossil data \cite{Alroy2008a}.
In contrast to ecological networks such as food webs, mutualistic networks and host-parasitoid networks \cite{Drossel2001a,Bascompte2007,Ings2009,Bascompte2009a}, the dependency network does not represent the interactions between individuals of different species but whether the existence of one species necessarily requires the presence of another species.

Two types of species are organized at certain dependency levels $l\ge 1$. Species at level $l=1$ are independent. In each iteration, 
they become extinct, with probability $\varepsilon$, or speciate to a new species, with probability $\mu$. 
Hence, the relative extinction probability
\begin{equation} \label{lambda}
\lambda = \varepsilon / \mu
\end{equation}
is the main parameter of the model.

In marine genera evidence has been reported for an age selectivity implying an extinction risk that ``drops off rapidly among the youngest age cohorts and thereafter shows little relationship to age'' \cite{Finnegan2008}.
Here we model speciation-extinction processes on long time scales.
Thus, firstly, we ignore the increased risk for the youngest cohorts and consider a constant extinction probability, known as Van-Valen's Law \cite{Valen1973}. 
Secondly, species at level $l\ge 2$ are directly dependent on only one other species \footnote{
If a species were dependent on the simultaneous presence of $m$ species and the extinction probability at level 1 is $\varepsilon_1$, species at higher levels would go extinct rapidly, because their extinction probability $\varepsilon_l = 1 - (1 - \varepsilon_{l-1})^m$ increases with $l$ and $m$.
If it were dependent on the presence of at least one of other $m$ species with extinction probabilities $\varepsilon$ it would show age selectivity because its extinction probability $\varepsilon_m(t) = \varepsilon^{m}$ increases while the species it depends on $E[m]=m_0 e^{-\varepsilon t}$ decreases with time.
}.
In each step, these species give rise to a new species with probability $\mu$, or become extinct if the species they depend on becomes extinct (Fig.\ \ref{fig:trees}).
Secondly, we ignore interspecies competition which may be a dominant force on short but not on long time scales \cite{Benton2009,Quental2013}.

The dependencies are determined by the following rules. 
Initially there are {$n_1(0)=k$} independent species at the lowest level $l=1$. If a species at level $l=1$ gives rise to a new species, then
 the new species is placed with a probability $\gamma\le 0.5$ at level $l=2$, thus being dependent, and with the probability $1-\gamma$
at level $l=1$, in this case being independent of other species.
If a species at a higher level $l\ge 2$ speciates, with the probability $\gamma$  we place the new species at level $l-1$, 
with probability $\gamma$ at level $l+1$, 
and with probability $1-2\gamma$ at level $l$, the level of its ancestor. 
When a species originates at level $l\ge 2$, it becomes dependent on a randomly chosen species located at level $l-1$.

\begin{figure}
    \includegraphics[width=8cm]{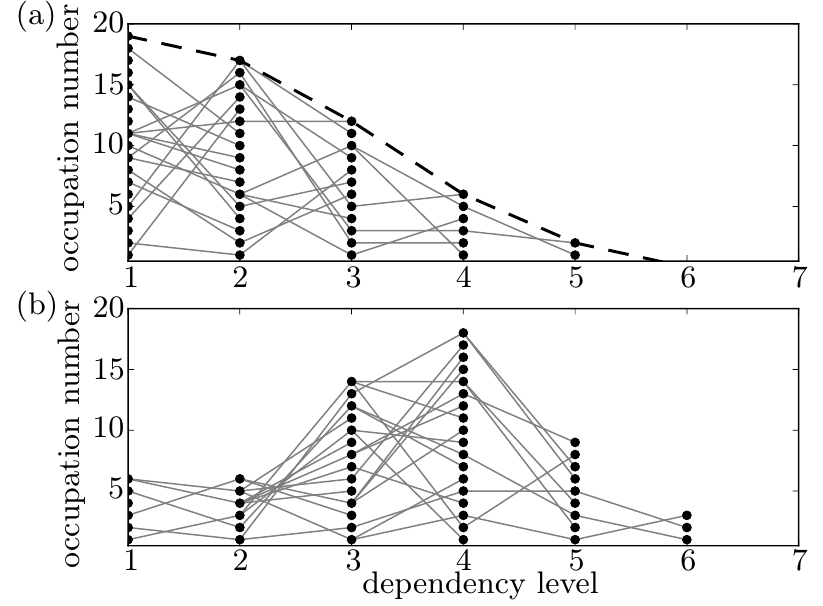}
    \caption{(a) Sketch of a dependency-network in the balanced state (a) and the imbalanced state (b). Both networks consist of the same number of species. The dashed line is close to a half-normal distribution.}
    \label{fig:cartoon-dependency-network}
\end{figure}

Using a mean field approximation for the dependency network and the continuum limit for time $t$, levels $l$ and occupation numbers $n_l$ we obtain the reaction-diffusion equation
\begin{equation}
\frac{\dif n_l}{\dif t} = \gamma\mu\frac{\dif^2 n_l}{\dif l^2} +  (\mu - \varepsilon) n_{l}, \ \ \text{with\ \ } \left. \frac{\dif n_l}{\dif l}\right|_{l=\frac{1}{2}}=0. \label{eqq:diffusion_equation}
\end{equation}
Since the model is defined only for $l\ge 1$, the Neumann boundary condition at $l=\frac{1}{2}$ ensures a zero net diffusion between $l=0$ and $l=1$. 
Regardless of the initial conditions, the occupation numbers $n_l$, the solution of Eq.\ (\ref{eqq:diffusion_equation}),
equilibrate to a half-normal distribution 
\begin{equation}\label{eq:dist}
n_l(t) = \frac{2 N(t)}{\sqrt{4 \pi\gamma\mu t}} \exp{\left( -\frac{(l-\frac{1}{2})^2}{4 \gamma\mu t} \right)},
\end{equation} 
see Fig.\ \ref{fig:cartoon-dependency-network}. 
The balanced state characterized by Eq.\ (\ref{eq:dist}) is the attractor of the dynamics, where the expectation value of the
extinction probability equals $\lambda N$ and the diversity grows exponentially, $N(t):=\sum_l n_l \sim e^{(\mu-\varepsilon)t}$. 

The lifetime distribution for species within a given time window of size $T$ follows an exponential decay 
\begin{equation}\label{eqq_L_T}
L_T(a)=\mu e^{-\mu a},
\end{equation}
where $a$ is the species age. This result is in agreement with the majority of the literature on marine species \cite{Sneppen1995a,Sole1996b,Newman1999b,Pigolotti2005}.

\paragraph{Episodic Stagnation} \label{sec:plateau}

All species which are dependent on a common species at level $1$ constitute a dependency tree. 
Given the survival of the root
the growth of tree
$i$ at level $l$ is governed by the simple differential equation
\begin{align}
\frac{\dif s_{il}}{\dif t} = \mu s_{i(l-1)}. \label{eqq:iterative_differential_equation} 
\end{align}
Given the root species of the tree
appears at time $t=t_i$, we have $s_{i1}(t)=1$ for $t\ge t_i $ which enables us to calculate the other occupation levels.
Specifically, for a relative extinction probability $\lambda\lesssim 1$ close to unity, 
 after a short transient period,
dependency trees  necessarily grow much faster than the (average) total diversity $N(t) \sim e^{(\mu-\varepsilon)t}$ as 
the solution of Eq.\ (\ref{eqq:iterative_differential_equation}) reads
\begin{align}
s_{il}(t) = \frac{1}{(l-1)!} (\mu(t-t_i))^{l-1} \ \ \text{for}\ t\ge t_i. \label{eqq:s_l}
\end{align}

A sum over all levels yields the size of the complete dependency tree
\begin{align}
S_i(t) = \sum_{l=1}^{\infty} s_{il} = \sum_{l=0}^{\infty} \frac{(\mu(t-t_i))^{l}}{l!} = e^{\mu(t-t_i)}. \label{eqq:total-tree-growth}
\end{align}
This means that the longer an independent species is spared from extinction, the more species are dependent on it and
that the number $S$ of species of a dependency tree increases exponentially.

Since the growth of a dependency tree becomes substantially accelerated
at higher levels (Eq.\ (\ref{eqq:s_l})),
a single tree may lead to a sudden imbalance of the entire dependency network such that many species at high levels depend on few independent species at level $l=1$ (Fig.\ \ref{fig:cartoon-dependency-network}).
In particular, when the dependency network has not returned to the balanced state large extinction cascades are more frequent.
During such time periods the temporarily increased extinction rate results in the suppression of the diversity growth and the emergence of a plateau.

\begin{figure}
    \includegraphics[width=8cm]{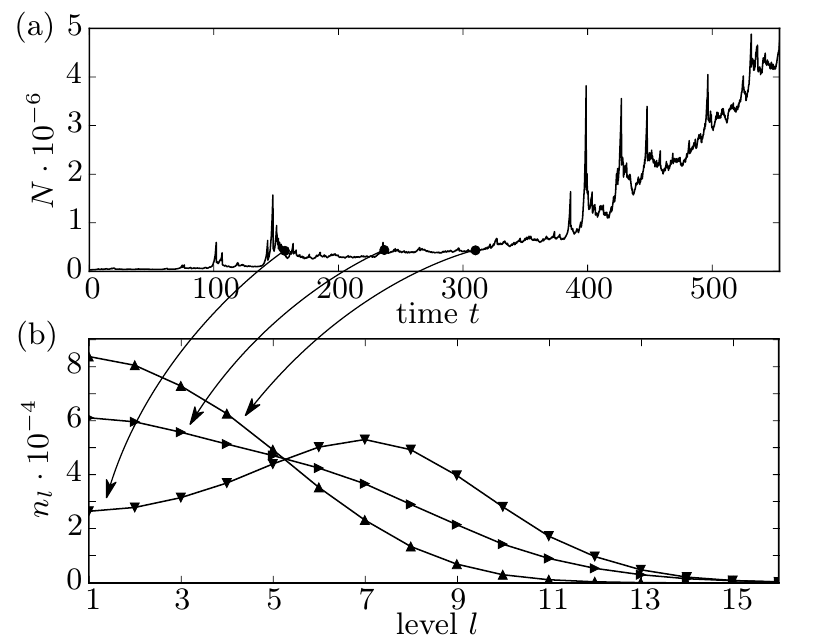}
    \caption{
Temporal development of a plateau.
Single realization for extinction probability $\lambda=0.985$ and $n_1(0)=1000$ initial species. 
(a) The total diversity curve with three indicated points of effectively constant diversity.
(b) The diversity distribution $n_l$ at the three indicated times in (a).
While the total diversity stagnates the dependency network reorganizes from the imbalanced state ($\blacktriangledown$) to a half-Gaussian distribution~($\blacktriangle$). 
}
    \label{fig:plateau}
\end{figure}

In Fig.\ \ref{fig:plateau} we have exemplified this behavior. 
An elevated value of the
diversity drives the system to an imbalanced state, and peaks in the diversity
are followed by relatively stable plateaus. 
Specifically, the system undergoes a transition back to a balanced state characterized by a half-normal distribution as depicted for three selected time points.

\begin{figure}
    \includegraphics[width=8cm]{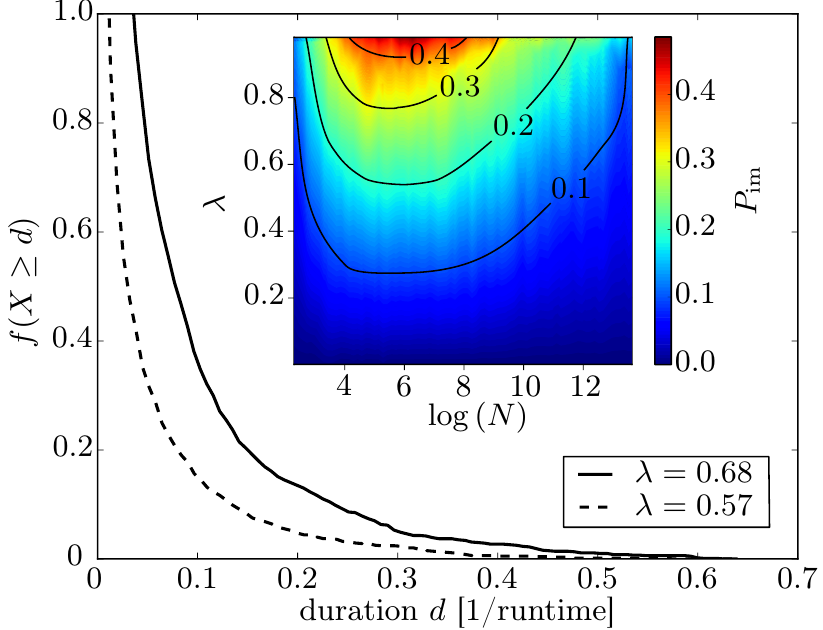}
    \caption{Frequency of periods of imbalanced states lasting longer than a duration $d$. Each realization was stopped at the time step when the diversity reached 1000 species. The inset shows the probability $P_\text{im}$ for an imbalanced state, based on $400$ realizations.}
    \label{fig:cdf-duration}
\end{figure}

The probability to find the system in an imbalanced state depends not only on $\lambda$ but also on the total diversity $N$. 
To examine this by computer simulations, we characterize an imbalanced state by the criterion $\text{argmax}_l(n_l)\neq 1$. Thus, the system switches to an imbalanced state as soon as the level $l=1$ ceases to be the most populated.
Fig.\ \ref{fig:cdf-duration} (inset) shows  
how the probability $P_\text{im}$ to find the dependency network in such an imbalanced state depends on $\lambda$ and $N$. 
The reason for this emerging pattern is that imbalanced states are caused by large dependency trees.  
This, however, becomes unlikely for $\lambda$ close to zero or a large total diversity.

\paragraph{Extinction rate distribution}

By employment of single event analysis \cite{Nagler2011,Nagler2012,Schroeder2013} we calculate the size distribution of extinction events, the number of species $S$ involved in single extinction cascades, which follows $P(S)=S^{-2}$ (\SI). Many models have primarily aimed at reproducing this power law behavior \cite{Newman1999,Amaral1999,Drossel2001}.
However, first, the distribution is not directly comparable to an extinction distribution obtained from the fossil data because the fossil data do not resolve distinct extinction cascades.
Second, the type of the extinction distribution in the fossil data is controversial \cite{Sole1997a, Newman1999a, Alroy2008a}. 
Since the extinction rate depends on the size of the considered time interval \cite{Camacho2000,Sole2002} and the total number of species, its fluctuations are only properly characterized by conditional probability measures. This suggests that a discrimination be made between the extinction rate and the size of an extinction event. 
For this reason we calculated in addition to the size distribution of extinction events the extinction rate distribution. Irrespectively of details of the extinction dynamics we analytically demonstrate that an exponentially increasing extinction rate (caused by an exponentially growing biodiversity) necessarily leads to an extinction rate distribution following a double power law with an exponent of $-1$ for small rates, and 
$-2$ for large rates, respectively (see \SI, Fig.~\ref{fig:size-distribution-num}).
Note that this prediction is exact.

\begin{figure}
    \includegraphics[width=8cm]{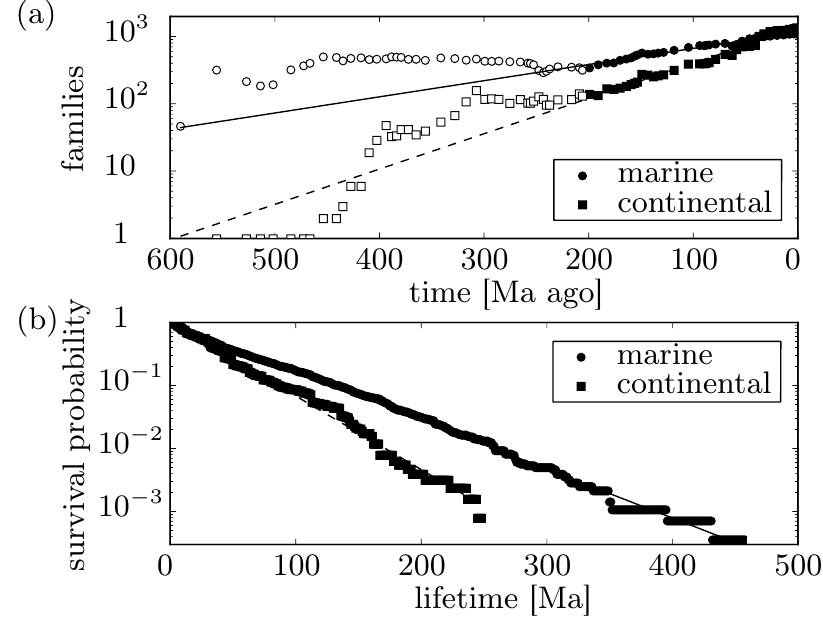}
    \caption{(a) Marine and continental diversity. 
The lines are fitted to the data in the time range \unit[-200]{Ma} to \unit[0]{Ma}, yielding ${\alpha=\unitfrac[0.0055(2)]{1}{Ma}}$ for marine diversity
 (${R=0.987}$, ${P<0.001}$), and ${\alpha=\unitfrac[0.0121(3)]{1}{Ma}}$ for continental diversity (${R=0.990}$, ${P<0.001}$).
(b) Survival probability (cumulative lifetime distribution) of marine and continental families.
Best fits are ${\mu=\unitfrac[0.01786(3)]{1}{Ma}}$ for marine data (${R=-0.998}$, ${P<0.001}$), and ${\mu=\unitfrac[0.0269(1)]{1}{Ma}}$ for continental data (${R=-0.990}$, ${P<0.001}$).}
    \label{fig:lambda-estimation}
\end{figure}

\begin{figure}
    \includegraphics[width=8cm]{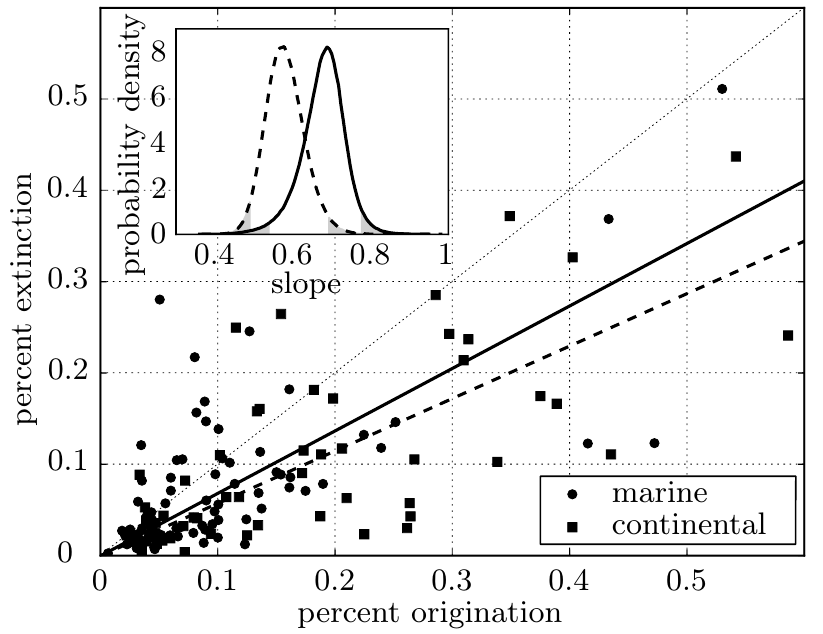}
    \caption{Percent extinction of total diversity against percent origination for each stage. The solid and the dashed line show the most frequent slope for marine and continental data found by least-squares-method in $10^6$ bootstrap samples. The inset shows the bootstrap distribution of the data. The gray areas are the 2.5\%-quantiles. The maxima are at slopes 0.683 (marine) and 0.567 (continental).
}
    \label{fig:lambda-estimation2}
\end{figure}

Next we ask whether and how marine and continental diversity are determined by different values of $\lambda$. We test this (on the level of families \footnote{We assume the large scale dynamics on the level of families to be similar to a coarse grained dynamics of the species level.}) by applying two different methods to the data of Fossil Record 2 \footnote{Although the Paleobiology Database offers continental data, too, the Fossil Record 2 is the established database for comparisons between marine and continental data.}.
Our model predicts the exponential growth of the diversity ${N(t) \sim e^{\alpha t}}$, ${\alpha:=\mu-\varepsilon}$,
together with the exponential decay of the lifetime distribution, Eq.\ (\ref{eqq_L_T}).
Fits of these functions to the fossil data yield the estimation of the parameters $\alpha$ and $\mu$, and thus an
estimation of the relative extinction probability ${\lambda = \frac{\mu-\alpha}{\mu}}$, Eq.\ (\ref{lambda}).
Applying this method to the fossil data (Fig.\ \ref{fig:lambda-estimation}), we obtain ${\lambda_\text{mar}=0.69(1)}$ for the marine, and ${\lambda_\text{cont}=0.55(2)}$ for the continental biodiversity.
In a second approach, independent of predictions of our model, we study the relation between percent extinction and percent origination
(Fig.\ \ref{fig:lambda-estimation2}).
Bootstrapping suggests that the slope for the marine data, $\lambda_\text{mar}$, is higher than the slope for the continental data, $\lambda_\text{cont}$ (evidence ratio of 9:1),
with the most likely values ${\lambda_\text{mar}=0.68}$ and 
${\lambda_\text{cont}=0.57}$.

Note that a conclusive comparison with the model is impossible because 
it would require multiple realizations of Earth's history.
Instead, we  ask how the behavior of the model qualitatively changes if $\lambda$
jumps from $\lambda_\text{cont}$ to $\lambda_\text{mar}$. 
As illustrated in Fig. \ref{fig:plateau} the duration of the imbalanced state correlates with the duration of the stagnation. 
Thus we infer from Fig. \ref{fig:cdf-duration} that stagnations lasting longer than a certain duration $d$ occur 
for $\lambda_\text{mar}=0.68$ more than twice as frequent as for $\lambda_\text{cont}=0.57$.

\paragraph{Discussion}

We have identified and studied a model where species speciate randomly and become extinct either due to endogenous or exogenous causes. Exogenous impacts occur with a constant probability whereas endogenous impacts are caused by extinction avalanches propagating through a system of interdependent species.

Assuming an expanding diversity, a long term stagnation, such as in the marine diversity, seems unlikely to be the coincidental result of exponential growth superimposed by random extinction events.
Using exact methods, we have demonstrated how a dependency network of species, which on very large scales grows exponentially, may evolve to an imbalanced state which implies long term stagnation.
Imbalanced states, however, are unstable and thus the network necessarily reorganizes to a balanced state and continues growing. 
This means that by taking the dependencies of species into account long stagnations turn out to be a typical behavior instead of a coincidental result, hence the imbalanced states are a plausible origin for long term stagnation of marine diversity.

The crucial parameter which determines this behavior is the ratio of the extinction to speciation probability. 
Two independent methods of analyzing the fossil data suggest that this ratio is substantially different for marine and continental diversity, which
provides a potential explanation for the qualitatively different growth of marine and continental diversity.


\clearpage
\onecolumngrid
\appendix

\renewcommand{\thepage}{S\arabic{page}} 
\renewcommand{\thesection}{S\arabic{section}}  
\renewcommand{\thetable}{S\arabic{table}}  
\renewcommand{\thefigure}{S\arabic{figure}}
\renewcommand{\theequation}{S\arabic{equation}}
\setcounter{page}{1}
\setcounter{section}{0}
\setcounter{table}{0}
\setcounter{figure}{0}
\setcounter{equation}{0}

\section{Supplementary Information}
\vspace{1.5cm}

\subsection{Marine and Continental Biodiversity}
\begin{figure}[h]
    \centering
    \includegraphics{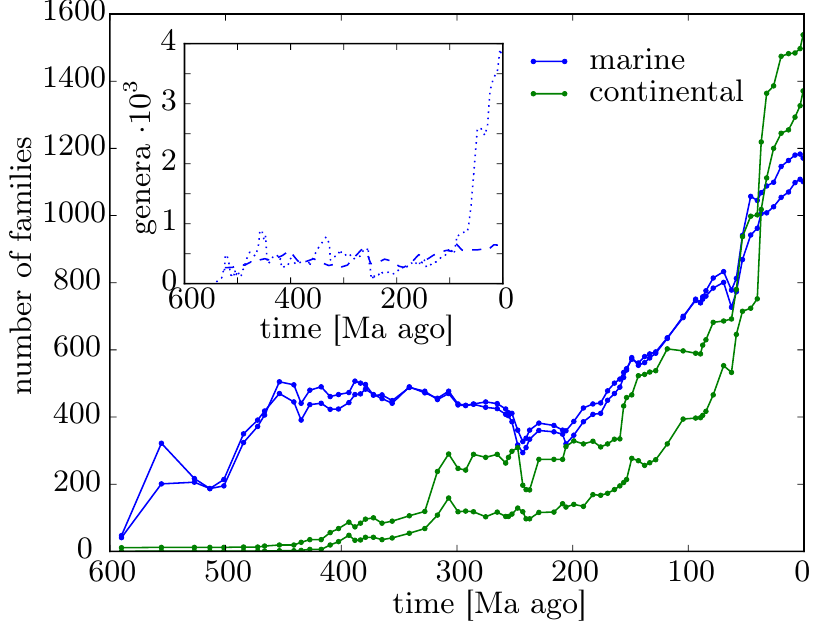}
    \caption{Evolution of biodiversity during the Phanerozoic on the level of families according to Fossil Record 2 \protect\cite{S:Benton1993}.
Two estimates are shown for the number of families for marine (blue) and continental (green) life, respectively. The inset shows the raw data of Sepkoski's compendium \cite{S:Sepkoski2002} (dotted line) and the sampling-standardized curve (dashed line) based on the Paleobiology Database \protect\cite{S:Alroy2008}, both for marine genera.}
    \label{fig:diversity-curve}
\end{figure}

Figure \ref{fig:diversity-curve} shows the biodiversity for marine and continental species from Ref.\ \cite{S:Benton1993}.
The inset displays two other diversity curves for marine species, but at a different taxonomic level (genera) and based on different databases and methods 
(raw data of Sepkoski's compendium \cite{S:Sepkoski2002} and sampling-standardized data of the Paleobiology Database \cite{S:Alroy2008}).
This demonstrates that arguments and empirical evidence in the debate on equilibrium and expansion are controversially.
In the following we derive an analytical framework that provides a possible explanation for marine and continental biodiversity growth.

\subsection{Level distribution of diversity}
We derive the level distribution  by solving a rate equation based on the rules of the model as described in the paper. If a level $l$ is populated with $n_l$ species, $\varepsilon n_l$ species will become extinct and $\mu n_l$ new species will be produced in one time step. A small ratio $\gamma$ of the new species will live at either a higher or a lower level. An exception is the lowest level $l=1$, where a ratio $\frac{\gamma}{2}$ of the new species live at level $l=2$ and the remaining new species stay at level $l=1$. This dynamics is described by the following rate equation: 
\begin{align}
\frac{\dif n_l}{\dif t} &= \mu \left( \gamma n_{l-1} + (1-2\gamma)n_{l} + \gamma n_{l+1} \right) - \varepsilon n_{l},\ \ \mathrm{for\ }l>1 \label{eq:dgl_nl} \\
\frac{\dif n_1}{\dif t} &= \mu (1-\gamma)n_{1} + \mu \gamma n_{2} - \varepsilon n_{1} \label{eq:boundarycondition}
\end{align}
We reduce equation \ref{eq:dgl_nl} by one parameter with a dimensionless time, $t\rightarrow \mu t$, and with a dimensionless parameter for a relative extinction probability $\lambda=\frac{\varepsilon}{\mu}$. By adding (virtually) $n_0$ as a boundary condition $n_0=n_1$, equation \ref{eq:dgl_nl} becomes valid for $l\ge1$ and equation \ref{eq:boundarycondition} is consequently redundant.
\begin{align}
\frac{\dif n_l}{\dif t} = &\gamma \left(n_{l-1}-2n_l+n_{l+1}\right) + (1-\lambda)n_l, \ \ \text{for }l\ge 1, \label{eq:discrete_diffusion} \\
&\text{with } n_0=n_1 \ \text{and } n_1(0)=k.
\end{align} 
We derive the growth of the total diversity by summation of equation \ref{eq:discrete_diffusion} over $l$: 
\begin{align}
\sum_{l=1}^{\infty} \frac{\dif n_l}{\dif t} &= (1-\lambda)\sum_{l=1}^{\infty}n_l + \gamma \sum_{l=1}^{\infty} \left(n_{l-1}-2n_l+n_{l+1}\right),\\
                      \frac{\dif N}{\dif t} &= (1-\lambda)N,\ \ \text{and thus} \label{eq:differential_growth}\\
                                    N(t) &= k e^{(1-\lambda)t}. \label{eq:growth}
\end{align}
To derive the distribution $n_l$ of the diversity among the levels we solve equation \ref{eq:discrete_diffusion}, which is apart from the term $(1-\lambda)n_l$ similar to the discrete diffusion equation (solved for example in~\cite{S:Lindeberg1990}). By using the ansatz $n_l(t) = e^{\alpha t}I_l(\beta t)$, where $I$ is the modified Bessel function of the first kind, and using the following relation for the derivative of $I_\eta(z)$ \citep[\S10.29.1]{S:Nist:Dlmf},
\begin{align}
2\frac{\partial}{\partial z}I_\eta(z) = I_{\eta-1}(z) + I_{\eta+1}(z), \label{eq:dBesselI_dt} 
\end{align}
we obtain a solution for equation \ref{eq:discrete_diffusion},
\begin{align}
n_l(t) = e^{(1-2\gamma-\lambda)t} I_l(2\gamma t).
\end{align}
To satisfy the boundary condition we add a virtual image of our solution at $l\le 0$ mirroring $l\ge 1$. The solution with arbitrary initial values $n_l(0)$ is
\begin{align}
n_l(t) = e^{(1-2\gamma-\lambda)t} \sum_{l'=1}^{\infty} n_{l'}(0) \left[ I_{l-l'}(2\gamma t) + I_{l-1+l'}(2\gamma t)\right]. \label{eq:general_solution}
\end{align} 
Since we restrict the initial condition to $k$ species at the lowest level and $0$ at higher levels, $n_l(0)=k \delta_{l-1}$, equation \ref{eq:general_solution} simplifies to
\begin{align}
n_l(t) = k e^{(1-2\gamma-\lambda)t} \left(I_{l-1}(2\gamma t)+I_{l}(2\gamma t)\right). \label{eq:bessel-distribution}
\end{align}
It will be useful to solve equation \ref{eq:discrete_diffusion} in a second way, which is a good approximation and provides a simpler expression for $n_l(t)$ than equation \ref{eq:bessel-distribution}. The difference equation \ref{eq:discrete_diffusion} becomes a partial differential equation by considering $l$ as a continuous variable, and replacing the second order difference quotient with respect to $l$ with the second derivative
\begin{align}
\frac{\dif^2 n_l}{\dif l^2} &= \frac{n_{l+\Delta l} - 2n_l + n_{l-\Delta l}}{\Delta l^2} + \mathcal{O}(\Delta l^2).
\end{align}
This yields a diffusion equation with a source term, whereby we replace the condition $n_0=n_1$ by a Neumann boundary condition at $l=\frac{1}{2}$.
\begin{align}
\frac{\dif n_l}{\dif t} = \gamma\frac{\dif^2 n_l}{\dif l^2} +  (1 - \lambda) n_{l}, \ \ \text{with\ \ } \left. \frac{\dif n_l}{\dif l}\right|_{l=\frac{1}{2}}=0 \label{eq:diffusion_equation}
\end{align}
As the initial value we choose $n_l(0)=2k\delta(l-\frac{1}{2})$, which is close to $n_1(0)=k$ but satisfies the Neumann boundary condition. \\
The solution is a Gaussian function with a growing height and a growing variance $\sigma_t^2 = 2\gamma t$,
\begin{align}
 n_l(t) &= \frac{2k}{\sqrt{4 \pi\gamma t}} \exp{\left( -\frac{(l-\frac{1}{2})^2}{4 \gamma t} + (1-\lambda)t \right)}
        = \frac{2 N(t)}{\sqrt{4 \pi\gamma t}} \exp{\left( -\frac{(l-\frac{1}{2})^2}{4 \gamma t} \right)}. \label{eq:gauss-distribution}
\end{align}

\subsection{Lifetime distribution}
As a first step, we derive the lifetime distribution $L'(a)$ of a fixed group of species with the relative extinction probability $\lambda$. This lifetime distribution is an exponential decay because the probability density $L'(a)$ to reach an age of $a$ decreases with the time by a factor of $\lambda$.   
\begin{align}
 \frac{\dif L'(a)}{\dif a} &= -\lambda  L'(a)  \\
  \Rightarrow L'(a) &= \lambda e^{-\lambda a} \label{eq:P_lifetime}
\end{align}
Recall that the lifetime distribution obtained from the fossil data is based on all species which became extinct during the Phanerozoic. We define a second lifetime distribution with a comparable definition. We call $L_T(a)$ the lifetime distribution of all species which emerged and died during a time interval $[0,T]$. This lifetime distribution is different from $L'(a)$ because new species originate during the time interval, thereby young species are more likely. \\
The number of species which originate at time $t$ is $N(t)=ke^{(1-\lambda)t}$. The probability density that one of these species, which originated during the time interval $[0,T]$, appeared at time $t$ is  
\begin{align}
P_T(t) &= \frac{N(t)}{\int_0^T N(t) \dif t} 
       = \frac{(1-\lambda)e^{(1-\lambda)t}}{(e^{(1-\lambda)T}-1)} 
       \approx (1-\lambda) e^{(1-\lambda)(t-T)},\ \ \text{for }T\gg 1.
\end{align}
The probability density to emerge at time $t$ and die at age $a$ is $P_T(t)L'(a)$. Now the lifetime distribution $L_T(a)$ is the integral of this product over all times $0\le t\le T-a$ (species, which appear later than $t=T-a$ and survive the time $a$ will not die within the interval $[0,T]$).
\begin{align}
L_T(a) &= c \int_0^{T-a} P_T(t)L'(a) \dif t
       = c \lambda \left( e^{-a}-e^{-\lambda a}e^{-(1-\lambda)T}\right) 
       \approx e^{-a} \label{eq:lifetime_probability}
\end{align}
The normalization constant is $c=\frac{1}{\lambda}$ and the approximation in the last step takes into account that $T\gg 1$. \\
Equation \ref{eq:lifetime_probability} written in terms of an age in real time instead of dimensionless time becomes $L_T(a)=\mu e^{-\mu a}$. Consequently, the lifetime distribution obtained from the fossil data with all species which became extinct during the Phanerozoic decays exponentially with the speciation probability (in contrast to the lifetime distribution $L'(a)$ written in terms of real time, which decays exponentially with the extinction probability).

\subsection{Extinction event distribution}
Since our model does not reach a stationary state, we define the extinction event distribution on a time interval $[0,T]$. The following calculation shows that this distribution follows a power law with exponent $-2$. \\ 
An extinction event is the coincident extinction of all species which belong to the same dependency tree. For their size distribution we need the lifetime of such a tree, which equals the already known lifetime distribution of species, and we need the size of a tree after a certain lifetime. \\ 
We call $s_{il}$ the number of species, which are dependent on the root species $i$ and live at level $l$, and $S_i=\sum_l s_{il}$ the size of the corresponding tree. Since every species at level $l=1$ is the root of its own tree, $s_{i0}=1$. The total number of species living at level $l$ is $n_l=\sum_i s_{il}$ and the total number of species in the system is $N=\sum_l n_l$. \\
In each time step $n_l$ new species are produced at level $l$ ($\mu n_l$ if we would not use dimensionless time) and to a good approximation this is the number of species placed at level $l$ (the errors of this approximation at all levels sum up to zero because the diffusion process is conservative).\\  
Each of these new species will be dependent on one species at the level below, thus only a fraction of $\frac{s_{i(l-1)}}{n_{(l-1)}}$ will become a part of tree $i$. Hence, the growth of $s_{il}$ is
\begin{align}
\frac{\dif s_{il}}{\dif t} =  \frac{s_{i(l-1)}}{n_{(l-1)}} n_l = f(l,t) \cdot s_{i(l-1)}, \label{eq:dsdt-first}
\end{align}
where $f(l,t):= \frac{n_l}{n_{(l-1)}}$. By substituting $n_l$ with equation \ref{eq:gauss-distribution} we get an expression for $f(l,t)$,
\begin{align}
f(l,t) = \frac{n_l}{n_{(l-1)}} = \exp{\left( \frac{1-l}{2\gamma t} \right)}. 
\end{align}
Recall that the level distribution $n_l(t)$ follows a Gaussian distribution with a variance $\sigma_t^2=2\gamma t$ (compare equation \ref{eq:gauss-distribution}). We approximate $f(l,t)$ by ignoring high levels where $n_l(t)\approx 0$ and including a constant fraction of species located at levels within $\epsilon\sigma$ standard deviations. The long-term limit of $f(\epsilon\sigma,t)$ is 
\begin{align}
\lim_{t\to\infty} f(\epsilon\sigma,t) &= \lim_{t\to\infty} \exp{\left(\frac{2-\epsilon\sqrt{2t}}{t} \right)} = 1, 
\end{align}
which means that we can assume $f(l,t) = 1$ for all $l \le \epsilon\sigma$.
Using this approximation, equation \ref{eq:dsdt-first} reduces to a simple differential equation. 
\begin{align}
\frac{\dif s_{il}}{\dif t} = s_{i(l-1)} \label{eq:iterative_differential_equation} 
\end{align}
If species $i$ appears at time $t=t_i$, we have $s_{i1}(t)=1$ for $t\ge t_i $ and we can calculate the other levels with a solution of equation \ref{eq:iterative_differential_equation},
\begin{align}
s_{il}(t) = \frac{1}{(l-1)!} (t-t_i)^{l-1} \ \ \text{for}\ t\ge t_i. \label{eq:s_l}
\end{align}
A sum over all levels yields the size of the complete dependency tree
\begin{align}
S_i(t) = \sum_{l=1}^{\infty} s_{il} = \sum_{l=0}^{\infty} \frac{(t-t_i)^{l}}{l!} = e^{t-t_i}. \label{eq:total-tree-growth}
\end{align}
The lifetime distribution of a dependency tree $S_i$ is the same as the lifetime distribution of species $L_T(a)=e^{-a}$ (equation \ref{eq:lifetime_probability}). The probability to die at an age $a=t-t_i$ within the interval $A$ should equal the probability to have a size within the size interval $S(A)$ at the moment of death.
\begin{align}
\int_A L_T(a) \dif a \stackrel{\text{!}}= \int_{S(A)} P(S) \dif S = \int_A P(S(a)) \left|\frac{\dif S}{\dif a}\right| \dif a 
\end{align}
The probability density $P(S)$ is the size distribution of the extinction events 
\begin{align}
P(S) &= L_T(a) \left(\frac{\dif S}{\dif a}\right)^{-1} = S^{-2},\ \ \text{for }S\ge 1. \label{eq:extinction-size-distribution}
\end{align}

\subsection{Extinction rate distribution in systems with exponentially increasing extinction rates}

In addition to the extinction size $S$ as the number of species that go extinct in a single {\em extinction event} we define the rate $r$ as the number of species that go extinct within $m$ {\em extinction events} occurring in a unit time step.
As we assume the number of extinction events $m$ to grow on average exponentially, $\langle m(t)\rangle=\alpha e^{\beta t}$, $\alpha$, $\beta$ being positive constants,
the probability $p(m)$ for any
series of extinctions whose total number add up to $m$ is given by 
\begin{equation}\label{eq:pm}
p(m) \approx \int \delta(m-\frac{\text{d}}{\text{d}t}\langle m(t) \rangle)\text{d}t = \int \delta(m-\alpha\beta e^{\beta t})\text{d}t \sim m^{-1}.
\end{equation}
We represent the size vector of
a series of $m$ extinction events with sizes $(S_1,S_2,...,S_m)$ by 
$\vec{S}\in \mathbb N^m$ ($\mathbb N$ are the natural numbers).
This series contains $r=\|\vec{S}\|_1=\sum_{i=1}^m S_i$ species and, if the $S_i$ are uncorrelated and $P(S)\sim S^{-2}$, occurs with probability 
$q_m(\vec{S})\sim \prod_{i=1}^{m} S_i^{-2}.$
The probability for the extinction of $r$ species, in a series of $m$ events within a unit time step, 
is given by the sum over all $m$-dimensional vectors of norm $r$,
\begin{align}  \label{eq:P_h(S)}
q(r|m) \sim \sum_{\vec{S}\in\mathbf{Q}_{mr}} \prod_{i=1}^{h} S_i^{-2} 
\end{align}
where $\mathbf{Q}_{mr} = \{\vec{S}\in \mathbb N^m|r=\|\vec{S}\|_1\}$ is the boundary of an $m$-dimensional simplex.

A proper way to discriminate
between the extinction size $S$ and the extinction rate $r$ is 
the conditional extinction rate distribution, obtained by combining Eq.\ (\ref{eq:pm}) and Eq.\ (\ref{eq:P_h(S)}),
\begin{align}  \label{eq:G_M}
D_M(r) \sim \sum_{m=1}^{m_\text{max}} p(m)q(r|m)  = \sum_{m=1}^{m_\text{max}} \frac{1}{m} \sum_{\mathbf{Q}_{mr}} \prod_{i=1}^{m} S_i^{-2}.
\end{align}
Eq.\ (\ref{eq:G_M}) describes the probability density for the integer extinction rate $r$, in any time interval with at most $M$ extinction events per time step.
The upper limit $m_{\text{max}}=\min{\{r,M\}}$ guarantees the summation over non-zero probabilities.

\begin{figure}
    \includegraphics{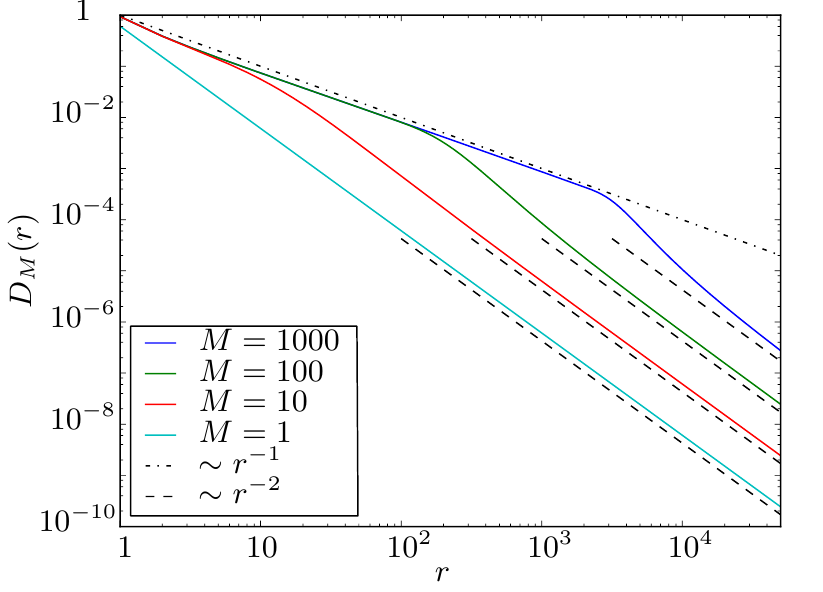}
    \caption{Conditional extinction rate distribution $D_M(r)$.
The solid curves show the distribution for different upper limits $M$ for the number of extinction events.
The dashed lines are power laws with exponents $-1$ and $-2$, respectively, according to the approximations of $D_M(r)$ for small, $r\lesssim M$,
 and for large extinction rates $r\gtrsim M$.}
    \label{fig:size-distribution-num}
\end{figure}

This relation allows us to prove that
in any ecosystem where the number of extinction events increases exponentially and 
the size of single extinction events is distributed as $q(S|1)=:P(S)\sim S^{-2}$,
fluctuations in the extinction rate $r$ are characterized by a bimodal extinction rate distribution, as shown in fig.\ \ref{fig:size-distribution-num}. Next we show that for $r\ll M$ the extinction rate distribution follows the power law $D_M(r)\sim r^{-1}$, where above $r\gtrsim M$ there is a transition to $D_M(r)\sim r^{-2}$.

\subsection{Approximations of the extinction rate distribution}
To expand the extinction rate distribution $D_M(r)$ for small and for large rates $r$ we
solve $q_m(\vec{S})$ for extrema on $\vec{S}\in\mathbf{Q}_{mr}$ by using Lagrange multipliers as
\begin{align}
\Lambda(\vec{S}) &= \prod_{i=1}^m S_i^{-2} + \lambda' (r - \|\vec{S}\|_1) \\
\frac{\partial }{\partial S_i}\Lambda &\stackrel{\text{!}}= 0 \ \ \ \Rightarrow \ \ \ \vec{S} = \sum_{i=1}^m \vec{e}_i \frac{2}{\lambda'} \prod_{j=1}^m S_j^{-2} = \sum_{i=1}^h \frac{r}{m}\vec{e}_i, \label{eq:min_vector}
\end{align} 
where $\{\vec{e}_1,...,\vec{e}_r\}$ is the standard basis of $\mathbb N^m$. 
Equation \ref{eq:min_vector} shows that the minimum of $q_m(\vec{S})$ on $\mathbf{Q}_{mr}$ is a vector in which all components have the same value \footnote{Actually, the $S_i$ are natural numbers and $S$ is not always divisible by $m$ without remainder. In this case, the minima are at the vectors close to the minimum vector in real space.}. 
The probability $q_m(\vec{S})$ increases with the distance (1-norm) of $\vec{S}$ to the minimum vector and reaches its maximum at the extreme points of $\mathbf{Q}_{mr}$, the vectors in which all components are $1$ except for one component, which gathers the remaining species $(r-m+1)$.\\
We define two subsets of $\mathbf{Q}_{mr}=\mathbf{Q}^{\text{ext}}_{mr} \cup \mathbf{Q}^{\text{res}}_{mr}$, where $\mathbf{Q}^{\text{ext}}_{mr}:= \{\vec{S}\in\mathbf{Q}_{mr}|q_m(\vec{S})=(r-m+1)^{-2}\}$ contains the extreme points of $\mathbf{Q}_{mr}$, which are the maxima of $q_m(\vec{S})$, and $\mathbf{Q}^{\text{res}}_{mr}=\mathbf{Q}_{mr}\setminus \mathbf{Q}^{\text{ext}}_{mr}$, which contains the residual vectors. \\
For the two cases $r\ll M$ and $r\gg M$ we calculate the extinction rate distribution $D_M(r)$ by taking into account the extreme points $\mathbf{Q}^\text{ext}_{mr}$ and regarding the residual vectors $\mathbf{Q}^{\text{res}}_{mr}$ as negligible.

\paragraph{Case $r\ll M$} The outer sum of $D_M(r)$ ends at $m_\text{max}=r$. If $m=r$, there is only one vector $\vec{S}=(1,1,...,1)$ in $\mathbf{Q}_{mr}$ with $q_m(\vec{S})=1$. If $m\le r$, $\mathbf{Q}^\text{ext}_{mr}$ has $m$ vectors with $q_m(\vec{S})=(r-m+1)^{-2}$.
\begin{align}
D_M(r) &\sim \sum_{m=1}^{m_\text{max}} \frac{1}{m} \sum_{\vec{S}\in\mathbf{Q}^\text{ext}_{mr}} \prod_{i=1}^{m} S_i^{-2} \nonumber\\
        &= \frac{1}{r} + \sum_{m=1}^{r-1} \frac{1}{m} \cdot \frac{m}{(r-m+1)^2} \nonumber\\
        &= \frac{1}{r} - 1 + \sum_{j=1}^r \frac{1}{j^2} = \frac{1}{r} - 1 + \text{H}_r^{(2)},
\end{align}
where $\text{H}_r^{(2)}\simeq \frac{\pi^2}{6}$ is the generalized harmonic number of order~$2$.

\paragraph{Case $r\gg M$} Here, $m_\text{max}$ is independent of $r$ and the outer sum ends at $m_\text{max}=M$. 
Again, $\mathbf{Q}^\text{ext}_{mr}$ contains $m$ vectors with $q_m(\vec{S})=(r-m+1)^{-2}$.
\begin{align}
D_M(r) &\sim \sum_{m=1}^{m_\text{max}} \frac{1}{m} \sum_{\vec{S}\in\mathbf{Q}^\text{ext}_{mr}} \prod_{i=1}^{m} r_i^{-2} \nonumber\\
        &= \sum_{m=1}^{M} \frac{1}{m} \frac{m}{(r-m+1)^2} \nonumber\\
        &= M r^{-2} + \mathcal O(r^{-3})
\end{align}

\subsection{Efficient numerical evaluation of the extinction rate distribution}
The direct evaluation of the extinction rate distribution,
\begin{align}
D_M(r) \sim  \sum_{m=1}^{m_\text{max}} p(m)q(r|m)  = \sum_{m=1}^{m_\text{max}} \frac{1}{m} \sum_{\mathbf{Q}_{mr}} \prod_{i=1}^{m} S_i^{-2},
\end{align}
for example to produce a plot of $D_M(r)$ as seen in figure \ref{fig:size-distribution-num}, is very time-consuming for large rates $r$ because $\mathbf{Q}_{mr}$ may contain a large number of vectors. Here we show how to calculate $q(r|m)$ as a recursive expression, which does not require to find the vectors of $\mathbf{Q}_{mr}$. 
The probability for a sequence represented by the vector $\vec{S}=(S_1,S_2,...,S_m)$ is equal to the product of the probabilities of the two vector decomposition $\vec{S'}=(S_1,S_2,...,S_{m-1})$ and $\vec{S''}=(S_m)$.
In this way we can express $q(r|m)$ as the Cauchy product of $q(r|m-1)$ and $q(r|1)$,
\begin{equation}
 q(r|m) = \begin{cases} \sum_{i=1}^{r-1}\limits q(i|m-1)q(r-i|1) &\text{if } r\ge m \\ 0 &\text{if }r<m \end{cases} \label{eq:P_h-recursive}.
\end{equation}
To compute $q(r|m)$ for all $r$ within the scope of interest we need to compute only a first sequence $(q(r|1))_r=(q(1|1),q(2|1),q(3|1),...)$ and repeatedly apply a convolution algorithm to get the succeeding sequence 
\begin{align}
(q(r|m))_r = (q(r|m-1))_r * (q(r|1))_r.
\end{align}

\end{document}